\documentstyle[12pt]{article}
\renewcommand{\baselinestretch}{2.0}
\begin{document}
\title{\large \bf Vortex Line Fluctuations in Model High Temperature
Superconductors}
\renewcommand{\baselinestretch}{1.0}
\author{\normalsize Ying-Hong Li \\
        \normalsize Institute for Theoretical Physics\\
        \normalsize University of Utrecht\\
        \normalsize 3508 TA Utrecht, The Netherlands \\
        \normalsize $\,$\\
        \normalsize and\\
        \normalsize $\,$\\
        \normalsize S. Teitel \\
        \normalsize Department of Physics and Astronomy\\
        \normalsize University of Rochester\\
        \normalsize Rochester, New York 14627}
\baselineskip=24pt
\date{\small (\today)}

\maketitle
\renewcommand{\baselinestretch}{2.0}
\small
\begin{center} Abstract \end{center}
We carry out Monte Carlo simulations of the uniformly frustrated
three dimensional
XY model, as a model for vortex line fluctuations in a high $T_c$
superconductor
in an external magnetic field.
A density of vortex lines of $f=1/25$ is considered.  We find two
sharp phase transitions.  The low $T$ superconducting
phase is an ordered vortex line lattice.
The high $T$ normal phase is a vortex line liquid, with much
entangling, cutting, and
loop excitations. An intermediate phase is
found, which is characterized as
a vortex line liquid of disentangled, approximately
straight, lines.  In this
phase, the system displays superconducting properties is the direction
parallel to the magnetic field, but normal behavior in planes
perpendicular to the field.  A detailed analysis of the vortex
structure function
is carried out.

\normalsize

\newpage

\noindent{\bf I.Introduction.}

While an understanding of
the microscopic mechanism leading to high temperature
superconductivity has remained elusive, much recent research has
focused on obtaining a phenomenological understanding of
the behavior of these materials in the presence of applied currents
and magnetic fields.  Such treatments have been based on the
familiar Landau-Ginzburg approach for type II superconductors,
which assumes only that macroscopically,
the superconductivity is adequately
described by a complex order parameter, as is the case for classical
BCS materials.  In this context,
the high $T_c$ materials are generally believed to
differ from the classical superconductors, in that fluctuation
effects are dramatically enhanced\cite{NelSeu,Nel,Fish}.  The
relatively high values of $T_c$, anisotropy, and ratio of magnetic
penetration to coherence lengths, $\kappa=\lambda/\xi_0$, in these
materials lead to
a significantly larger critical region in temperature, about $T_c$.

In the presence of an applied magnetic field $H_{c1}<H<H_{c2}$,
the low temperature phase
of a pure system is the Abrikosov vortex
line lattice.  As temperature
increases, it has been argued that fluctuations will cause
this vortex line lattice to
melt at a temperature well below that
given by the mean field $H_{c2}(T)$
line, leading to a vortex line liquid
phase\cite{NelSeu,Nel,Fish,Hou,Glaz}.
Experimental evidence for such a scenario has been cited
in mechanical oscillator measurements\cite{Gam}, as well as by the
observation that the onset of sizable reversible magnetization
generally occurs at
a temperature well above that where resistivity vanishes\cite{Wor},
an indication of strong fluctuation effects.
Several theoretical works have sought to describe the nature of this
vortex line lattice melting transition, as well as predict properties
of the vortex line liquid phase.
One commonly used approach is an effective elastic
theory\cite{NelSeu,Fish,Hou,Glaz},
in which the energy of transverse fluctuations of the vortex
lines from their lattice positions, is computed in terms of non-local
elastic constants determined from the mean field Landau-Ginzburg
theory\cite{Bra5}.
The Lindemann criterion is then used to estimate melting, when such
transverse fluctuations become a sizable fraction of the vortex line
spacing.  Another theoretical model has been the ``two dimensional
boson" approximation\cite{NelSeu,Nel,NelDou},
in which vortex lines, fluctuating in a directed
fashion along the direction of the applied magnetic field,
are viewed as the ``world lines" of two dimensional interacting bosons.
Many-body methods applied to the two dimensional boson problem have
then been used to estimate melting in the dilute
line limit\cite{NelSeu,Nel},
as well as to compute properties of the vortex line
liquid\cite{NelSeu,NelDou}.

The above theoretical models suffer from several important limitations.
The effective elastic theory loses validity once the vortex lines are
in a liquid phase.  The ``$2d$ boson" approximation
assumes a simplified vortex line interaction, that applies mainly
when lines do not wander too much from parallel alignment with
the applied magnetic field.
Both models assume that the vortex lines thread the system
in a directed fashion; each line passes through
each plane perpendicular to $\vec B$ only once.  Closed vortex
line loop excitations, which have been argued to play a significant
role in the $\vec B=0$ case\cite{Fish,Minn}, as well as in the finite
$\vec B$ case\cite{Carn}, are ignored.  In view of this, it
is useful to have a well defined statistical model, for which
numerical simulations can be carried out, and in which
the effects of vortex line fluctuations may be quantitatively examined.

In this paper we present the results of such a study, based on
a simplified model which we believe captures the essential physics
of a fluctuating Landau-Ginzburg superconductor; the uniformly
frustrated three dimensional XY model.  We have previously introduced
this model in two earlier works\cite{LiTei1,LiTei2}.
Here we provide details, as well
as surprising new results.  We find that as the vortex line lattice
is heated, the system undergoes two distinct thermodynamic
transitions.  First there is a transition characterized by
the loss of phase coherence in planes perpendicular to the applied
magnetic field, while phase coherence is retained parallel to the
field.  In this intermediate region,
the vortex line lattice disorders in the
transverse direction, while individual lines remain essentially
straight and unentangled from each other.  Then at a higher
temperature, phase coherence is lost in the direction parallel to
$\vec B$ as well.  This high temperature
region is characterized by vortex line
wandering and many closed loop excitations.  The vortex lines are
highly entangled, yet they easily cut through one another.

The remainder of our paper is organized as follows.  In section II
we present a detailed description of our model, and the approximations
involved.  In section III we present the results of our Monte
Carlo simulations:  section IIIA discusses computation of the
helicity moduli, which measure the loss of phase coherence;  section
IIIB discusses the nature of vortex line fluctuations in the
various thermodynamic phases; and section IIIC presents a detailed
analysis of the vortex structure function, which is used to extract
effective elastic constants and correlation lengths.  In section IV
we summarize our conclusions.

\noindent {\bf II. Description of the Model}

It is generally believed that the phenomenological behavior
of the high temperature superconductors is well described by
the familiar Landau-Ginzburg free energy functional\cite{Tink},
\begin{equation}
F[\psi (\vec r),\vec A (\vec r)]=\int d^3r
\left\{ \alpha |\psi |^2
+{1\over 2}\beta |\psi |^4 + {\hbar^2\over 2m}|(-i\vec\nabla +
{2\pi\over\Phi_0}\vec A)\psi |^2 + {1\over 8\pi}
|\vec\nabla\times\vec A |^2\right\}\label{eq:Fcont}
\end{equation}
where $\psi$ is the superconducting order parameter, $\vec A$ the
magnetic vector potential, and $\Phi_0=hc/2e$ the flux quantum.
The local magnetic induction is
$\vec b(\vec r)=\vec\nabla\times\vec A(\vec r)$.
The first two terms represent the condensation energy, the third
term is the kinetic energy of the flowing supercurrents, and the
fourth term is the magnetic field energy.  We choose to work
throughout this
paper with the $Helmholtz$ free energy, which is a function of the
average magnetic induction, $\vec B =(1/Vol)\int d^3r
\vec b(\vec r)$,
inside the bulk superconductor.  The applied magnetic
field $\vec H$, may in principle be determined by an appropriate
thermodynamic derivative of the Helmholtz free energy,
but its value is of no direct
concern for our calculations.  For simplicity,
we consider here only a material with isotropic couplings, and indicate later
how the effects of anisotropy may be included in our model.

The mean field solution, appropriate for low temperature
superconductors, is obtained by minimizing Eq.(\ref{eq:Fcont}) with
respect to both $\psi$ and $\vec A$, subject to the constraint
of constant $\vec B$.  For $\vec B=0$, one gets the uniform solution
$\psi(\vec r)=\sqrt{|\alpha|/\beta}\equiv \psi_0$
in the superconducting phase,
$\alpha <0$ ($\psi_0=0$ in the normal phase, $\alpha>0$).
For $0<B<H$, in the mixed phase of a type II
superconductor, the result is the familiar Abrikosov
flux tube lattice\cite{Tink}.
Each flux tube carries one $\Phi_0$ of flux, and
consists of a vortex line in the phase $\theta (\vec r)$ of
$\psi (\vec r)=|\psi (\vec r)|{\rm e}^{i\theta (\vec r)}$.  The vortex
line is surrounded by a normal
region of radius the coherence length,
\begin{equation}
\xi_0=\hbar/\sqrt {2m|\alpha |}\label{eq:xi}
\end{equation}
and the magnetic field is confined to within a region of radius
the magnetic penetration length,
\begin{equation}
\lambda = \sqrt{ mc^2/16\pi e^2|\psi_0|^2}\label{eq:lam}
\end{equation}
The density of flux tubes is $B/\Phi_0$, hence
the separation between vortex lines is \begin{equation}
a_v\simeq \sqrt{\Phi_0/B}\label{eq:av}
\end{equation}
For the high temperature superconductors, we now wish to compute
thermodynamic quantities, averaging over all fluctuations of $\psi$
and $\vec A$ (subject to the constraint of constant $\vec B$),
weighting configurations with the Boltzman factor
${\rm e}^{-F[\psi,\vec A]/k_BT}$.  In order to carry out this average
numerically, with Monte Carlo simulations, we make the following
approximations.

For $\lambda\gg a_v$, the magnetic fields of the individual flux tubes
strongly overlap, leading to an approximately uniform local magnetic
induction.
In this limit, we therefore make the approximation
of replacing the fluctuating $\vec A$ with a fixed one, giving a
spatially uniform magnetic induction $\vec B = \vec\nabla\times\vec A$.
We will discuss the consequences of this approximation on the
effective vortex line interaction in more detail below.
Next we discretize\cite{latgrad} our model by placing it on a
periodic cubic mesh
of lattice constant $a$,
resulting in the free energy functional
\begin{equation}
F[\psi_j,A_{ij}]=a^3\sum_j
\left\{ (\alpha +{\hbar^2\over 2ma^2})|\psi_j|^2+
{1\over 2}\beta |\psi_j|^4\right\} -
{a\hbar^2\over 2m}\sum_{\langle ij\rangle} \psi_j^*{\rm e}^{-iA_{ij}}
\psi_i + c.c.\label{eq:Fdis}
\end{equation}
where we have dropped the now constant magnetic energy term.  Here
$i,$ $j$ denote sites of the cubic mesh,
${\langle ij\rangle}$ denote nearest neighbor bonds of the
mesh, and
\begin{equation}
A_{ij}\equiv {2\pi\over\Phi_0}\int_i^j\vec A\cdot d\vec l\label{eq:Aij}
\end{equation}
Finally, we choose the lattice constant of the numerical mesh $a=\xi_0$,
the coherence length of Eq.(\ref{eq:xi}).  Thus the decay to zero of
the superconducting wavefunction $\psi$ at the normal core
of a vortex line, takes place within one unit cell of the mesh.
Outside the normal core, we can make a London
approximation and set $|\psi|=\psi_0$, a constant.  Thus on the mesh, we
consider only fluctuations of the form
$\psi_j=\psi_0{\rm e}^{i\theta_j}$,
where the amplitude is fixed, and only the phases $\theta_j$ vary.
Details concerning the normal cores of vortex lines
are treated approximately by the
small length cutoff $a=\xi_0$ imposed by the mesh.  Applying this to
the free energy in Eq.(\ref{eq:Fdis}) results, within additive
constants, in the effective Hamiltonian for a
frustrated three dimensional XY model,
\begin{equation}
F[\theta_i]=-J_0\sum_{\langle ij\rangle}\cos (\theta_i-\theta_j-A_{ij})
\label{eq:Ham}
\end{equation}
where Eqs.(\ref{eq:lam},\ref{eq:Fdis}), $\Phi_0=hc/2e$, and
$\xi_0=a$ give,
\begin{equation}
J_0={\Phi_0^2\xi_0\over 16\pi^3\lambda^2}\label{eq:J}
\end{equation}
Anisotropy, or impurities, could now easily be
incorporated into the model,
by letting the coupling $J_0$ vary appropriately
on different bonds of the
mesh.  In this work we continue to consider only
the case of isotropic couplings.
We expect that the $qualitative$ behaviors we find, will be to a large
extent the same as in anisotropic systems\cite{LiTei1,Blat}.

The Hamiltonian (\ref{eq:Ham}) above has been studied extensively
in two dimensions as a model for Josephson junction
arrays\cite{TeiJay,JJs}, and
in two and three dimensions as a model for
granular superconductors\cite{Stro}.  Its application to behavior
in the high temperature superconductors, is best realized in the limit
$\xi_0\ll a_v\ll \lambda$, where the above approximations should apply.
This condition may be rewritten\cite{Tink} as
$H_{c1}\ll H\ll H_{c2}$.  For the
high $T_c$ materials, where $\kappa=\lambda/\xi_0\sim 100$
is large, this will
cover a large range of experimentally accessible magnetic fields $H$.
For YBaCuO, for example, this condition holds for fields on the
order of a few Tesla.

Much of the theoretical work concerning fluctuations in the high $T_c$
superconductors, has been discussed in terms of
fluctuations of the vortex
lines induced by a finite $B$.  It is therefore of interest to consider
the effective vortex line interaction in our
model (\ref{eq:Ham}).  Towards
this end, it is convenient to replace the cosine function
of Eq.(\ref{eq:Ham})
with the periodic Gaussian, or Villain function\cite{Vill}.
\begin{equation}
F[\theta_i]  =  J_0\sum_{\langle ij\rangle}
V(\theta_i-\theta_j-A_{ij})\label{eq:Hvil}
\end{equation}
where
\begin{equation}
V(\alpha)  \equiv  -(T/J_0) \ln \{ \sum_{m=-\infty}^\infty \exp[-
{1\over 2}J_0(\alpha -2\pi m)^2/T]\} \label{eq:vil}
\end{equation}
This has the effect of eliminating
the coupling between spin wave and vortex excitations of the
phase variables $\theta_i$ in
the Hamiltonian (\ref{eq:Ham}).

Making a standard duality transformation\cite{Frad}
on this new Hamiltonian (\ref{eq:Hvil}), gives the effective
Hamiltonian in terms of the vortex degrees of freedom.
\begin{equation}
F_v[n_\mu({\rm i})]=2\pi^2 J_0\sum_{\rm i,j}
\sum_{\mu}[n_\mu({\rm i})-f_{\mu}({\rm i})]
[n_\mu({\rm j})-f_{\mu}({\rm j})]
G(\vec r_{\rm i}-\vec r_{\rm j})\label{eq:vline}
\end{equation}
Here $n_\mu({\rm i})$ is the integer vorticity in the phase $\theta$,
going around the plaquette with normal
$\hat
\mu=\hat x$, $\hat y$, $\hat z$ of the cubic unit cell centered at
the dual mesh
site i, and $f_\mu ({\rm i})$ is the number of flux quanta of magnetic
induction through this plaquette,
\begin{equation}
2\pi f_\mu({\rm i}) ={2\pi\over\Phi_0}\oint \vec A\cdot d\vec
l= A_{jk}+A_{kl}+A_{lm}+A_{mj}\label{eq:f}
\end{equation}
where the sum is taken counterclockwise
over the bonds forming the edges of the plaquette.
For our uniform magnetic field approximation, with $\vec B$ along the
$\hat z$ direction,
\begin{equation}
f_z({\rm i})=B\xi_0^2/\Phi_0\equiv f,
\qquad f_x({\rm i})=f_y({\rm i})=0\label{eq:f2}
\end{equation}
In this case, the ground state of Eq.(\ref{eq:vline}) will consist of
a periodic structure of straight vortex lines, aligned with $\vec B$
in the $\hat z$ direction, with density $f$.

The interaction potential $G$ in Eq.(\ref{eq:vline}),
is the lattice Green's function which solves
\begin{equation}
D_{\rm ij} G(\vec r_{\rm j}-\vec r_{\rm k})=-\delta_{\rm i,k}
\label{eq:gdef}
\end{equation}
where $D_{\rm ij}$
is the lattice Laplacian.
$G(\vec r)$ is most easily expressed in terms of its Fourier transform
\begin{equation}
G(\vec r)={1\over N}\sum_k G_k{\rm e}^{i\vec k\cdot\vec r}\label{eq:Gft}
\end{equation}
where $N$ is the number of sites in the mesh, and
\begin{eqnarray}
G_k & = & 1/K^2=1/(K_z^2+K_\perp^2),\nonumber\\
K_z^2 & \equiv & 2-2\cos
k_za,\qquad K_\perp^2 \equiv 4-2\cos k_xa - 2\cos
k_ya\label{eq:Green}
\end{eqnarray}
Note $K^2\simeq (ka)^2$ for small $ka$, however they are different
at large
$ka$ due to the discreteness of the mesh. For large $r\gg a$,
$G(\vec r)\simeq 1/(4\pi r)$.

We thus see that the interaction between vortex lines
in our model is the Coulomb interaction, $G_k=1/K^2$.
For comparison, the
mean field London interaction in the continuum\cite{Bra2},
obtained by minimizing the free
energy Eq.(\ref{eq:Fcont}) with respect to a spatially varying
$\vec b(\vec r)$, is
$G_k^{London}=1/(k^2 +\lambda^{-2})$, which is a screened
Coulomb interaction with screening length $\lambda$.
Thus our approximation of taking a uniform magnetic induction
$\vec b(\vec r)=\vec B$,
and ignoring
spatial variations and fluctuations in $\vec b$, corresponds
formally to making this
screening length infinite.  Our model therefore strongly suppresses long
length scale (small $k$) vortex line density fluctuations, as
compared to the true
superconductor.  However we expect that the energetics of
fluctuations on small length scales
$k^{-1}\mathrel{\hbox{\hbox to 0pt
{\lower.5ex\hbox{$\sim$}\hss}\raise.4ex\hbox{$<$}}}
\lambda$ are well approximated.  Provided $a_v\ll\lambda$, the
model should
therefore correctly describe the high $k\sim 1/a_v$
fluctuations describing
collisions between near neighbor vortex lines, which
are believed responsible
for the melting of the vortex line lattice\cite{Hou}.

It will be convenient for later to write Eq.(\ref{eq:vline}) in terms
of Fourier transforms.  Defining
\begin{equation}
n_\mu (\vec k)\equiv\sum_{\rm i} n_\mu({\rm i})
{\rm e}^{-i\vec k\cdot \vec r_{\rm i}}
\label{eq:nk}
\end{equation}
we have as the Hamiltonian
\begin{equation}
F_v[n_\mu({\rm i})]={2\pi^2 J_0\over N}
\sum_{\mu,k\ne 0}n_\mu(\vec k)n_\mu(-\vec k)
G_k\label{eq:vlinek}
\end{equation}
The $k=0$ contribution to $F_v$ would be
$[n_\mu(\vec k=0)-Nf_\mu]^2 G_{k=0}$,
where $f_\mu$ is as in Eq.(\ref{eq:f2}).
However as $G_{k=0}$
diverges (see Eq.(\ref{eq:Green})), the requirement that
the total free energy be
finite, insures that we only have configurations where
the average vortex density
in direction $\hat\mu$ is $f_\mu$; hence the $k=0$ piece vanishes.

\noindent {\bf III. Monte Carlo Simulations}

To investigate the behavior of the model, we carry out Metropolis Monte
Carlo simulations, of the Hamiltonian (\ref{eq:Hvil}),
on a cubic mesh with
$N=L_zL_\perp^2$ sites and periodic boundary conditions
in all directions.
Our simulations are carried out directly in terms of the phase
variables $\theta_i$,
with no additional presumptions concerning the behavior of vortex lines,
beyond those implicit in Eq.(\ref{eq:Hvil}).
Simulations of this same model, only working directing in
terms of the vortex variables and using the Hamiltonian (\ref{eq:vline}),
have recently been reported by
Cavalcanti $et$ $al$\cite{Braz}.

Previously we have reported\cite{LiTei1,LiTei2} on
our simulations using a density of field induced vortex lines of
$f=1/5$.  In this work we report on simulations using a density of lines
$f=1/25$, with an average spacing between lines of $a_v=5a$.
This value of $f$ was chosen so as to increase the ratio
$a_v/a$, and  hence reduce effects
due to the discreteness of our numerical mesh, while making $f$ large
enough so as to have a large number of lines contained within a system
of computational size.  Our present simulations are carried out for
a mesh of size $L_\perp=50$ and $L_z=24$, giving $100$ vortex lines in
the ground state. The ground state of our system, we believe to be the
periodic extension of the unit cell of
straight vortex lines, shown in Fig.1.  Although we
have no way to prove that this is the true ground state, anealing with
Monte Carlo simulations failed to result in any state with lower energy.
That our presumed ground state is a square lattice of lines, instead of
the triangular lattice expected in a continuum model, is due to the
competition between vortex line interactions and the cubic
discreteness of our numerical mesh.
Most of our calculations represent runs of $2,000$ Monte Carlo sweeps to
equilibrate, with $20,000$ sweeps for computing averages.  Each Monte
Carlo sweep refers to one pass through the entire cubic mesh.
Errors are estimated either by block averaging, or from comparing
independent runs.
Henceforth we measure length in units
where  $\xi_0=a=1$, and temperature in units where $k_B=1$.

{\bf A. Helicity Modulus}

The most convenient method to detect phase coherence,
ie. superconductivity,
in the model Hamiltonian (\ref{eq:Hvil}), is to compute the helicity
modulus\cite{JasFis}.
The helicity modulus $\Upsilon_\mu$ in the direction $\hat \mu$,
is defined as follows\cite{TeiJay2}.
Imagine applying boundary conditions such that the phase across
the system in the direction $\hat\mu$ is twisted by an amount
$L_\mu\delta $.  If $\hat\mu=\hat x$ for example, we would require,
\begin{equation}
\theta (x=L_x,y,z)-\theta(x=0,y,z)=L_x \delta
\label{eq:tbc}
\end{equation}
Now transform to the new set of variables
\begin{equation}
\theta^\prime_i=\theta_i-(\vec r_i\cdot\hat\mu)\delta
\label{eq:tprime}
\end{equation}
If the $\theta_i$ obey the twisted boundary conditions (\ref{eq:tbc}),
then these new variables $\theta^\prime_i$ are easily seen to obey
periodic boundary conditions.  In terms of the
$\theta^\prime_i$, the system
with the twisted boundary conditions is described by the Hamiltonian
\begin{equation}
F[\theta_i^\prime;\delta]=J_0\sum_{\langle ij\rangle}
V(\theta_i^\prime-\theta_j^\prime-A_{ij}-
(\hat e_{ij}\cdot\hat\mu)\delta)\label{eq:Fd}
\end{equation}
where $\hat e_{ij}$ is the unit vector from site $i$ to site $j$.
One can now form the partition function
\begin{equation}
Z(\delta)=\sum_{\{ \theta^\prime_i\} }
{\rm e}^{-F[\theta_i^\prime;\delta]/T}\label{eq:Zd}
\end{equation}
where the sum is only over
configurations $\{\theta_i^\prime\}$ satisfying
periodic boundary conditions.  The total free energy is then
\begin {equation}
{\cal F}(\delta)=-T\ln Z(\delta)\label{eq:FTd}
\end{equation}
and the helicity modulus is now defined as
\begin{equation}
\Upsilon_\mu\equiv {1\over N}{\partial^2 {\cal F}\over\partial\delta^2}
\Big|_{\delta=0}\label{eq:upsi}
\end{equation}
where $N=L_zL_\perp^2$.
$\Upsilon_\mu$ measures the stiffness to twisting the phase up
in the $\hat\mu$ direction, and may be thought of as the renormalized
coupling $J_0$ on infinitely
long length scales.  Equivalently, $\Upsilon$ may be viewed
as the effective
superfluid density, or via Eq.(\ref{eq:J}), as determining the
renormalized magnetic penetration length, $\Upsilon\sim 1/\lambda^2_R$.
When $\Upsilon_\mu>0$ there is phase coherence in the
$\hat\mu$ direction, and when $\Upsilon_\mu=0$, phase coherence is lost.
The vanishing of $\Upsilon_\mu$ in a particular direction therefore
signals a superconducting to normal transition.

In terms of the definitions Eq.(\ref{eq:Fd}$-$\ref{eq:upsi}), we can
write the helicity modulus as the correlation function,
\begin{eqnarray}
& &\Upsilon_\mu  =  {J_0\over N}\left\langle
\sum_{\langle ij\rangle}V^{\prime\prime}
(\theta_i-\theta_j-A_{ij})(\hat e_{ij}
\cdot\hat\mu)^2\right\rangle\label{eq:helmod}\\
& - &{J_0^2\over TN}\left\{ \left\langle
\left[ \sum_{\langle ij\rangle}
V^\prime(\theta_i-\theta_j-A_{ij})
(\hat e_{ij}\cdot\hat\mu)\right]^2\right\rangle
-\left\langle\sum_{\langle ij\rangle}
V^\prime(\theta_i-\theta_j-A_{ij})
(\hat e_{ij}\cdot\hat\mu)\right\rangle^2\right\}
\nonumber
\end{eqnarray}
which can be evaluated directly in the
Monte Carlo simulation.  $V^\prime$
and $V^{\prime\prime}$ are the first and second derivatives of
the Villain function defined in Eq.(\ref{eq:vil}).

In Fig. 2 we show our results for $\Upsilon_z$ and $\Upsilon_\perp
\equiv{1\over 2}(\Upsilon_x+\Upsilon_y)$, as a function of temperature.
Both heating
from the ground state (Fig.1), and cooling from random are shown.
The results clearly indicate two sharp transitions.  As $T$ is increased,
there is first a transition $T_{c\perp}\simeq 1.3J_0$ at which
$\Upsilon_\perp\to 0$ but $\Upsilon_z>0$.
Phase coherence is lost in the directions perpendicular to the applied
magnetic field, however phase coherence persists in the direction
parallel to the magnetic field.
Then, at a higher $T_{cz}\simeq 2.5J_0$, $\Upsilon_z\to 0$ and phase
coherence is lost in the parallel direction as well, and we have the
completely normal phase.
In the intermediate phase, $T_{c\perp}<T<T_{cz}$,
the linear resistivity $\rho_z$ parallel to
$\vec B$ should vanish, however
the linear resistivity $\rho_\perp$ perpendicular to $\vec B$ should
remain finite.
Such an intermediate phase is in complete
agreement with recent predictions
by Feig'lman and co-workers\cite{Fei}, who used arguments based
on the ``$2d$ boson" approximation.
At the lower transition, $T_{c\perp}$, a slight hysteresis is seen,
comparing heating versus cooling.  No hysteresis is seen at $T_{cz}$.

{\bf B. Vortex Line Fluctuations}

We now seek to describe the nature of vortex line fluctuations in the
three distinct phases discussed above.
We will find that the high
temperature phase is one in which vortex lines wander, entangle, and
cut through each other easily.  In this phase, we also find that the
density of vortex lines is greatly increased by the excitation of many
closed vortex loops.
The intermediate phase is one in which
vortex lines are unentangled, remaining on average straight and aligned
with $\vec B$, however they are in a liquid state as far as their
correlations in planes perpendicular to $\vec B$ are concerned.

As our simulations are carried out in terms of the variables $\theta_i$,
we first must locate the positions of the vortex lines as they thread
their way through the system.  We use the following algorithm.  For
each plaquette $\alpha$ in the system, we compute
\begin{equation}
 2\pi f_{\alpha} + \sum_{\alpha} [\theta_i-\theta_j-A_{ij}]
\equiv 2\pi n_\alpha\label{eq:vortcy}
\end{equation}
where the sum is taken counterclockwise around
the bounds of the plaquette, $f_\alpha$ is the
magnetic flux through the plaquette as given by
Eqs.(\ref{eq:f},\ref{eq:f2}),
and the $[\theta_i-\theta_j-A_{ij}]$ are measured restricted to the
interval $[-\pi,\pi]$.
$n_\alpha$ is then the integer vorticity
in the $\theta_i$ going around this plaquette.  As vorticity is
conserved, each unit cell of the cubic mesh which has one plaquette with
a vortex entering the cell, has another plaquette with a vortex leaving.
By computing the vorticity entering and
leaving the unit cells of the mesh,
we are able to trace out the paths of the vortex lines.  In the event
that two vortex lines enter and leave the same unit cell,
we randomly assign
which exiting segment is connected to which entering segment, for the
purpose of identifying the paths of those lines.

{\bf (i) Upper Transition $T_{cz}$}

One property that has received much theoretical attention, has been
the entanglement of vortex lines, once the ground state vortex line
lattice has melted.  Interesting dynamical effects have been suggested
assuming a large energy barrier exists against vortex lines cutting
through each other\cite{NelSeu,Nel,MarNel1,ObuRub}.
In such a case, entanglement would provide a topological constraint
limiting vortex line motion.
A single pinned line would pin the many
lines with which it is entangled,
substantially reducing the flux flow resistance.

In our model, it is simple to check for entanglement.
We follow a prescription proposed by Nelson\cite{NelSeu,Nel}.
Let $\{ \vec r_{\perp i}(0), \vec r_{\perp j}(0), . . .\}$
be the positions
of the vortex lines in the $xy$ plane, as they pierce the plane at
$z=0$.  Let
$\{\vec r_{\perp i}(L_z), \vec r_{\perp j}(L_z), . . .\}$
be the corresponding locations of
these vortex lines as they pierce the $z=L_z$ plane.
Since our simulation uses periodic boundary
conditions in the $z$ direction, parallel to $\vec B$,
all the $\vec r_{\perp i}(L_z)$ must connect onto some
$\vec r_{\perp j}(0)$,
so that the vortex lines remain continuous in the periodically extended
system.  If $\vec r_{\perp i}(L_z)$ connects to $\vec r_{\perp i}(0)$,
we say that the line  $i$ reconnects to itself.  If we view the
periodic boundary condition along the $\hat z$ direction
as representing the circumference of a three dimensional
torus, then line
$i$ winds only once around the torus, before reconnecting to itself.
In contrast,
if $\vec r_{\perp i} (L_z)$ connects to
$\vec r_{\perp j}(0)$, $j\ne i$, then
lines $i$ and $j$ are part of a larger multiple connection of lines that
winds more than once around torus, before reconnecting
to itself (see Fig.3).
For an entangled vortex line state, we would expect there
to be many such
multiple connections, while for disentangled approximately straight
lines, there should be only simple self-reconnections.
In Fig. 2, we plot
the fraction of lines which reconnect to themselves, $R$,
versus temperature,
which we compute while cooling the system from an initial
random configuration.
$R$ drops sharply to zero at $T_{cz}$, from its low
temperature value of one.  We thus identify this
upper transition, where phase coherence in the direction
parallel to $\vec B$
is lost, as a transition from straight disentangled lines, to wandering
entangled lines.  Further, the fact that the multiply
connected entangled
lines present at high temperature, completely disentanlge to simple
self-reconnected lines below $T_{cz}$,
suggests that the lines may easily
cut through each other as they disentangle.

As an alternative criterion for vortex line entanlgement,
we can consider the transverse fluctuations
of the vortex lines $\langle
u^2(l_z)\rangle$ as a function of length
$0\le l_z\le L_z$ parallel to
$\vec B$,
\begin{equation}
\langle u^2(l_z)\rangle = {1\over N_v l_z}
\sum_{i=1}^{N_v}\sum_{z=0}^{l_z}\langle
[\vec r_{\perp i}(z)-\vec r_{\perp i}(z_0)]^2\rangle,
\qquad z_0=l_z/2 \label{eq:u2}
\end{equation}
where $\vec r_{\perp i}(z)$ is the position of line $i$
in the $xy$ plane at $z$,
$N_v=fL_\perp^2$ is the number of magnetic field induced lines, and
the sum is along
the lengths of these lines for a distance $l_z$.
Vortex lines in our model may in general consist
of two types: the magnetic
field induced lines which thread the
entire system from $z=0$ to $z=L_z$,
and thermally excited closed vortex loops.
In the evaluation of Eq.(\ref{eq:u2})
we do not include sums over vortex line segments which
belong to the thermal
loops.  In the event that a loop intersects a field induced line, we
randomly assign which line segment ``belongs" to the
field induced line, and
which segment ``belongs" to the thermally excited loop.

For straight, unentangled vortex lines,
we expect that $\langle u^2\rangle$
will saturate to a constant as $l_z$ gets large.
For wandering, entangled vortex
lines we expect\cite{NelSeu} that $\langle u^2\rangle$
will grow with increasing $l_z$ as,
\begin{equation}
\langle u^2(l_z)\rangle = u_0^2(T) + D(T)l_z\label{eq:u2fit}
\end{equation}
$D(T)$ measures the diffusion of the line
in the transverse direction, as
one moves down the $\hat z$ axis, and it can be used to determine the
``entanglement correlation length,"\cite{NelSeu}
\begin{equation}
\xi_z(T)\equiv {a_v^2\over 4 D(T)}\label{eq:xiz}
\end{equation}
$\xi_z$ measures the distance one has to move
in the $\hat z$ direction, in
order to have a transverse fluctuation equal to half the average spacing
between the vortex lines.  $\xi_z$ is the length scale on which the
fluctuating lines can approach each other, leading to a mutual winding and
hence entangling.

In Figs.4$a$,$b$ we plot $\langle u^2(l_z)\rangle$, versus $l_z$,
for several different values of temperature.
For $T\ge2.5J_0\simeq T_{cz}$ (Fig.4$a$),
the data points for the three largest values of $l_z$ show a clear
linear increase with $l_z$, as expected from Eq.(\ref{eq:u2fit})
for an entangled phase.
For $T<2.5J_0$ (Fig.4$b$), there remains a systematic curvature
in the data, suggesting a saturation of $\langle u^2\rangle$
 at large $l_z$, as in an unentangled phase; the data here is limited by
finite size effects,
and runs in a system with larger $L_z$ would be needed
in order to clarify the situation.
Obtaining $D(T)$ from a fit of Eq.(\ref{eq:u2fit}) to the three largest
values of $l_z$, for the data of Fig.4, we compute the resulting $\xi_z$
from Eq.(\ref{eq:xiz}) and show the result in Fig.5.

In a similar manner, we can compute the ``cutting correlation length,"
$\xi_c$, which we define as the average distance one has to
move in the $\hat z$
direction, between two successive cuttings of a given
vortex line.  Within
our model, we define a cut as whenever there
are two or more vortex lines
entering and leaving a given unit cell of the numerical mesh.
Such a situation
represents an overlapping of the vortex cores,
and there is no constraint
whatever on how the lines will subsequently fluctuate
out of this unit cell;
lines are free to cut through each other, or even to cut and reconnect
different segments of different lines.  If $N_c$ is
the average number of total cuttings observed in the simulation, then
\begin{equation}
\xi_c(T)\equiv {N_vL_z\over N_c(T)}\label{eq:xic}
\end{equation}
We plot $\xi_c(T)$ in Fig.5.

Comparing
the entanglement and cutting correlation
lengths, $\xi_z$ and $\xi_c$, we
see that for $T\ge 2.5J_0\simeq T_{cz}$, $\xi_z\simeq \xi_c$.
Thus, entanglement and cutting
are occurring on the same length scales for the region above $T_{cz}$,
and here we would expect entanglement to
provide little constraint on the motion of vortex lines.
Below $T_{cz}$, we see $\xi_c$ growing faster than $\xi_z$.
However, as discussed above in connection with Fig.4$b$, we believe that
our data here is limited by finite size effects, and hence the values
shown in Fig.5 are not a reliable indication of behavior in an
infinite sample.  It therefore remains unclear whether
there exists some narrow range of temperature, at or below $T_{cz}$, in
which entanglement might lead to constraints on vortex line motion.

Finally, as an additional measure
of the fluctuations of the vortex lines,
we calculate the average density of vortex lines cutting the
planes perpendicular
to $\vec B$,
\begin{equation}
n_v\equiv {1\over N}\sum_{\rm i}\langle n_z^2({\rm i})
\rangle\label{eq:nv2}
\end{equation}
where the sum is over all dual sites i,
and $n_z({\rm i})$ is the vorticity
passing through the plaquette in the $xy$ plane above site i.
In virtually all cases, $n_z=0,\pm 1$,
and so $n_v$ counts the density of
lines, irrespective of the sign of the vorticity.
If the only vortex lines
present in the system are the magnetic field induced lines,
and these lines
wander through the system in a directed fashion (ie. all lines pierce
each $z$ plane only once; there is no ``snaking back")
then one has $n_v=f$.  If however, there are additional
closed vortex line loops excited, or lines are not directed,
then $n_v$ will
be larger than $f$.  In Fig.6 we plot $n_v(T)$.  At low temperatures
$T\mathrel{\hbox{\hbox to 0pt
{\lower.5ex\hbox{$\sim$}\hss}\raise.4ex\hbox{$<$}}}
2.0J_0$, we see that $n_v\simeq f$, and hence
there are only directed
field induced lines.  However for $T
\mathrel{\hbox{\hbox to 0pt
{\lower.5ex\hbox{$\sim$}\hss}\raise.4ex\hbox{$>$}}} 2.0J_0$,
$n_v$ grows dramatically,
becoming $n_v\simeq 2f$ at the transition $T_{cz}$.
Thus, by the time the
transition is reached, there are many thermally
excited closed vortex loops in
the system.  This suggests that the transition at
$T_{cz}$ is mediated, not
by the wandering of the field induced vortex lines, but rather by the
proliferation of thermally excited vortex line loops.  The behavior of
transverse fluctuations and cuttings, shown in
Figs.4 and 5, may be understood
as arising from the growth of loop excitations,
that intersect and link up
neighboring field induced lines, rather than the direct wandering of the
field induced lines.  In fact, Fig.6 shows a qualitative resemblance to
the density of vortices as one crosses the transition, in the ordinary
three dimensional XY model\cite{3dxy}
(corresponding to $\vec B$=0).  Similarly, a plot
of the specific heat $C$, which we compute by numerical
differentiation of the
average energy and show in Fig.7, shows the same cusp-like feature at
$T_{cz}$ as in the $3d$ XY model\cite{LiTei3}.  We note that these loop
excitations, which we find important for the transition at $T_{cz}$,
are explicitly
excluded in the ``$2d$ boson" and effective elastic medium
approximations for vortex line fluctuations, as well as in other earlier
simulations\cite{Ryu}.

{\bf (ii) Lower Transition $T_{c\perp}$}

The low temperature phase $T<T_{c\perp}$ is characterized by an ordered
vortex line lattice, with periodicity as given in Fig.1.  The vanishing
of $\Upsilon_\perp$ above $T_{c\perp}$, together with the result that
$R\simeq 1$ for $T_{c\perp}<T<T_{cz}$ (see Fig.2), indicates that at    t
$T_{c\perp}$ the vortex line lattice melts into a liquid of unentangled
vortex lines.  One way to measure this transition, is by considering the
local orientational order of vortices in the planes perpendicular to
$\vec B$.  These vortices are just the intersections of the vortex lines,
with a particular plane at constant $z$.
We define the $n$-fold {\it local} orientational
order, in terms of the quantity
\begin{equation}
\Psi_n={1\over fN}\sum_{z,i}{1\over M_i}\sum_{j=1}^{M_i}
{\rm e}^{in\phi_{j,j+1}(i)}
\label{eq:orient}
\end{equation}
In the above, $z$ labels the planes perpendicular
to $\vec B$, $i$ labels the
vortices in the plane $z$, and $j$ labels the $ M_i$ nearest neighbor
vortices of $i$ in plane $z$. $\phi_{j,j+1}(i)$ is the angle between
the line from $i$ to $j$, and the line from $i$ to
$j+1$, which is the next neighbor after $j$ which one encounters in
rotating around $i$ (see Fig.8). The nearest neighbors of $i$ are
determined by the method of Voronoi polygons\cite{Vor}.
For perfect square orientational order,
with all $\phi_{j,j+1}(i)=\pi/2$,
such as in the ground state of
Fig.1, we have $\Psi_4=1$, while $\Psi_6={\rm e}^{i\pi}$.  For perfect
hexatic orientational order, with all $\phi_{j,j+1}(i)=\pi/3$,
$\Psi_6=1$, while $\Psi_4={\rm e}^{i4\pi/3}$.
In Figs.9$a$,$b$, we plot respectively the amplitude, and phase of
$\Psi_4$ and $\Psi_6$.  Both amplitudes show a clear kink at
$T\simeq 1.3J_0
\simeq T_{c\perp}$.  At this temperature, the phases indicate a clear
cross over from the $4$-fold order of the
ground state vortex line lattice,
to the local $6$-fold order of a vortex line liquid.

For a more global picture of the transition from vortex line lattice to
vortex line liquid, we have computed the correlation of vortices
in the planes perpendicular to $\vec B$.  If we define the vortex
structure function\cite{NelSeu,NelDou} as
\begin{equation}
S(\vec k_\perp,k_z)\equiv {1\over N}\langle
n_z(\vec k)n_z(-\vec k)\rangle
\label{eq:Sk}
\end{equation}
with $\vec k = (\vec k_\perp,k_z)$ and $n_z(\vec k)$
as in Eq.(\ref{eq:nk}),
then the correlation between vortices in different planes separated
by a height $z$, is
given by the Fourier transform of
Eq.(\ref{eq:Sk}) with respect to $k_z$,
\begin{equation}
S(\vec k_\perp,z)\equiv {1\over L_z}\sum_{k_z} S(\vec k_\perp,k_z)
{\rm e}^{ik_zz}
\label{eq:Skz}
\end{equation}
The correlation between vortices in the {\it same} plane is then just
$S(\vec k_\perp,z=0)$.  In Figs.10$a$-$d$ we show intensity plots of
$S(\vec k_\perp,z=0)$ for $k_x,k_y$ in the interval $[-\pi,\pi]$,
for several values of temperature, upon heating the system.  At
$T=0.6J_0<T_{c\perp}$ (Fig.10$a$)
we see sharp Bragg peaks, corresponding to the periodic
vortex line lattice of Fig.1, superimposed on a small
continuous background.
The width of these Bragg peaks, corresponds to the finite resolution
of wavevectors allowed by our finite system, $\Delta k_\mu =2\pi/L_\mu$.
At $T=1.5
\mathrel{\hbox{\hbox to 0pt
{\lower.5ex\hbox{$\sim$}\hss}\raise.4ex\hbox{$>$}}}
T_{c\perp}$ (Fig.10$b$),
we see an approximately rotationally invariant
structure, with peaks of finite width.
The first peak is at a radius $k_\perp=2\pi/a_v$,
corresponding to the average
separation between vortices; the smaller, second peak, is
at $k_\perp=4\pi/a_v$.  Such a structure function is
characteristic of a liquid.
The lack of complete rotational symmetry in the peaks
(aside from a slight 4-fold distortion to be expected from
the underlying cubic mesh) we believe is due to the
finite time duration of
our simulation; we see no evidence for long range orientational order
as might be present in a hexatic line liquid\cite{MarNel3}.
As $T$ increases (Fig.10$c$), the second peak disappears into the
continuous background.
At $T=2.5J_0\simeq T_{cz}$ (Fig.10$d$),
both peaks have virtually disappeared, and only a
continuous background remains.  Thus for $T
\mathrel{\hbox{\hbox to 0pt
{\lower.5ex\hbox{$\sim$}\hss}\raise.4ex\hbox{$>$}}} T_{cz}$,
even the short range order, characterized by the average spacing between
field induced vortex lines, is lost; this is consistent with the
proliferation of many vortex line loops, as evidenced by Fig.6.

In the intermediate phase, we find that the system relaxes
{\it very} slowly
to equilibrium.  We illustrate this by computing the time dependent
correlation,
\begin{equation}
g(t)={1\over N}\sum_{\rm i} \left\{
\langle n_z({\rm i},t)n_z({\rm i},0)\rangle -
\langle n_z({\rm i},t)\rangle \langle n_z({\rm i},0)\rangle \right\}
\label{eq:gt}
\end{equation}
where the time $t$ is just the number of Monte Carlo sweeps through the
entire cubic mesh.  We plot $g(t)$ at $T=1.9J_0$ in Fig.11.  After an
initial fast decay, we observe a slow algebraic decay,
$g(t)\sim t^{-\zeta}$
with $\zeta\simeq 0.23$.
We believe that it is this very slow decay which is responsible for the
small but finite values of $\Upsilon_\perp$ that are seen for
$T>T_{c\perp}$ in Fig.2, rather than usual finite size effects.
When we doubled
the number of Monte Carlo steps in our simulation, we observed
$\Upsilon_\perp$ in this region to decrease.

Because of this slow equilibration,
as the system is cooled back below $T_{c\perp}$, we have been unable
to regain the ordered vortex line lattice of Fig.1.  Instead, we cool
into a structure consisting of perfectly straight lines,
ordered into local
domains of the ground state, with frozen in domain walls. In Fig.12
we show an intensity plot of $S(\vec k_\perp,z=0)$ at $T=0.6J_0$, for
such a cooled configuration.
The two rings of peak intensity which exist for $T>T_{c\perp}$ (see
Fig.10$b$), are now narrow and
``squared off."
This feature of freezing in domain walls, as well as the
hysteresis in $\Upsilon_\perp$ seen in
Fig.2, suggest that the transition at $T_{c\perp}$ may be
first order\cite{Huse}.
Another possibility is that $T_{c\perp}$ is not a true thermodynamic
transition at all, but only a cross-over depinning temperature where
the effective energy barriers to vortex line motion grow so large,
that vortex motion is frozen out on the scales of our simulation.

{\bf C. Vortex Structure Function}

In Fig.10, we used the structure function $S(\vec k_\perp,k_z)$
to qualitatively
determine the nature of the vortex line structure in the
different thermodynamic
phases.  Here we present a more detailed analysis of
$S(\vec k_\perp,k_z)$,
from which we can extract information about the
effective elastic constants of the model.

Following the works of Marchetti\cite{Mar} and Nelson
and LeDoussal\cite{NelDou},
the natural form for the coarse-grain averaged
free energy of our model, valid on $hydrodynamic$ length scales
$k_\perp^{-1} >a_v$ in the vortex line liquid phase, is
\begin{equation}
F={1\over 2Nf^2}\sum_k\left( c_{44}(\vec k)\sum_{\mu_\perp  =x,y}
\left[\delta n_{\mu_\perp}(\vec k)\delta
n_{\mu_\perp}(-\vec k)\right] +
c_l(\vec k)\delta n_z(\vec k)\delta n_z(\vec k)\right)
\label{eq:Fgrain}
\end{equation}
where the $\delta n_\mu(\vec k)$
give the deviation from the average
uniform density $f_\mu$,
and are now viewed as independent continuous variables.
$c_{44}$ and
$c_l$ are the tilt and bulk moduli of the vortex lines on
the numerical mesh.

Using the constraint that vortex lines must be continuous,
ie. the discrete divergence of $n_\mu({\rm i})$ vanishes,
\begin{equation}
\sum_\mu [n_\mu ({\rm i}+
\hat\mu)-n_\mu ({\rm i})]=0,\qquad {\rm or}\qquad
\sum_k n_\mu (\vec k)[{\rm e}^{ik_\mu}-1]=0\label{eq:vlcont}
\end{equation}
one can do the appropriate Gaussian integrals\cite{NelDou}
over $\delta n_\mu(\vec k)$
and compute the structure function of Eq.(\ref{eq:Sk})
\begin{equation}
S(\vec k_\perp,k_z)={Tf^2K_\perp^2\over
c_l(\vec k)K_\perp^2+c_{44}(\vec k)
K_z^2}\label{eq:Shydro}
\end{equation}
where $K_\mu^2$ is related to $k_\mu$ as in Eq.(\ref{eq:Green}).
The factor $K_\perp^2$ in the numerator of the
expression above, determines the qualitative
form of the continuous
background of $S(\vec k_\perp,z=0)$ seen in Figs.10.

{}From Eq.(\ref{eq:Fgrain}) and the form of the bare vortex
line interaction given in
Eqs.(\ref{eq:Green},\ref{eq:vlinek}), we expect
\begin{equation}
c_l(\vec k)=4\pi^2 f^2J_1 G_k,\qquad\qquad
c_{44}(\vec k)=c_l(\vec k)+f\epsilon_1(k_z)\label{eq:emod}
\end{equation}
where we write $J_1$ instead of $J_0$
to allow for possible renormalization
of this coupling in the coarse-graining procedure.  $\epsilon_1$
is the single vortex line tension, which depends on $k_z$, due to the
interaction between the different segments of the single line.
The difference $c_{44}-c_l$ just gives the additional energy needed to
create the elongation of the vortex lines described by the transverse
components of the fluctuation\cite{Mar}. Substituting Eq.(\ref{eq:emod})
into Eq.({\ref{eq:Shydro}) gives,
\begin{equation}
TK_\perp^2/S(\vec k_\perp,k_z)=4\pi^2J_1 + [\epsilon_1(k_z)/f]K_z^2
\label{eq:tks}
\end{equation}

We now compute $S(\vec k_\perp,k_z)$
directly within the Monte Carlo simulations.  Calculations are done for
values of $k_z=2\pi m_z/L_z$, $m_z = 0,1,...,L_z/2$, and for the
special values  of $\vec k_\perp=k_\perp\hat x$
and $k_\perp\hat y$, where
$k_\perp=2\pi m_\perp /L_\perp$, $m_\perp =1,...,L_\perp/2$.
By symmetry, we
expect that $S(k_\perp\hat x,k_z)=S(k_\perp\hat y,k_z)$,
hence we combine our results by defining
$S(k_\perp,k_z)$ as the average of $S$ in these two special directions.

In Figs.13$a-f$, we plot $TK_\perp^2/S$,
as a function of $K_\perp^2$ and
$K_z^2$ for temperatures $T=1.0J_0$ (in the vortex line lattice phase),
$T=2.25J_0$ (in the intermediate phase), and $T=2.75J_0$ (in the normal
phase).  In the line lattice phase (Figs.13$a,b$),
we see that $TK_\perp^2/S$ is approximately
independent of $k_\perp$, obeying the hydrodynamic
form of Eq.(\ref{eq:tks})
at all $\vec k$.  In the intermediate, unentangled
line liquid phase (Figs.13$c,d$),
a strong $k_\perp$ dependence appears, with a minimum appearing at
$K_\perp^2 \simeq 1.2$, very close to the value of $1.38$ which
corresponds to $k_\perp = 2\pi/a_v$.  In the
normal phase (Figs.13$e,f$) as $T$ increases,
this minimum decreases to smaller $k_\perp$,
reflecting the increasing density of vortex lines (see Fig.6; this
also explains why the minimum in Fig.13$c$ is not $exactly$ given
by $k_\perp = 2\pi/a_v$).
This minimum as a function of $k_\perp$, is a reflection of the peak in
$S(\vec k_\perp,k_z)$ which occurs isotropically in the
$\vec k_\perp$ plane at
$|\vec k_\perp|\simeq 2\pi/a_v$, in the line liquid phase. It may be
viewed as an effective softening of the bulk modulus
$c_{44}(\vec k)$ at these
large values of $k_\perp$.  In the line lattice phase,
this isotropic peak
contracts into discrete Bragg peaks,
none of which lie along the $\hat x$
or $\hat y$ directions (see Fig.10$a$). Hence the $k_\perp$ dependence
of $TK_\perp^2/S$ flattens out.

Although Eqs.(\ref{eq:Shydro}-\ref{eq:tks}) were proposed for the
line liquid phase, similar results have been given for the line lattice
phase, using an effective elastic approximation\cite{NelDou}.
In this case the only
change is that one must replace $c_{l}$ in Eq.(\ref{eq:Shydro})
with $c_{11}\equiv c_{l}+c_{66}$, where $c_{66}$ is the shear modulus.
This results
in the addition of the term $c_{66}K_\perp^2$ to Eq.(\ref{eq:tks}).
It is unclear whether or not the differences observed in the
$k_\perp$ dependence of Figs.13$a$ and $c$,
may be interpreted in terms of the vanishing
of the shear modulus\cite{MarNel3}
at the melting transition $T_{c\perp}$.

Comparing the curves of $TK_\perp^2/S$ versus $K_z^2$,
for different $k_\perp$,
in Figs.13$d$ and $f$, one sees that
the dependence of $TK_\perp^2/S$ on $k_\perp$ and
$k_z$ cannot be separated out into the simple form
$4\pi^2J_1(k_\perp) + [\epsilon_1(k_z)/f]K_z^2$
as might be expected from the ``$2d$ boson" approximation\cite{NelDou}.
As we have noted in earlier work\cite{LiTei2},
the more complex dependence
we observe, may
be interpreted as adding a $\vec k_\perp$ dependence to the line
tension $\epsilon_1$.  Using Eq.(\ref{eq:tks}) to define the
effective line tension
as $k_\perp\to 0$,
\begin{equation}
\epsilon_1(k_z)\equiv {f\over K_z^2}\left[ {TK_\perp^2\over S(0,k_z)}-
{TK_\perp^2\over S(0,0)}
\right]\label{eq:eps1}
\end{equation}
we plot in Fig.14 $\epsilon_1(k_z)$ versus $\ln K_z^2$ for
several temperatures $T$.
The plots approach straight lines at large $K_z^2$.
This is in qualitative
agreement with the expression for the line tension obtained from an
elastic approximation, expanding about the line lattice in a layered
superconductor\cite{Glaz},
\begin{equation}
\epsilon_1(k_z)={\Phi_0^2\over 2(4\pi\lambda)^2}
\ln\left[ {k_{\perp,max}^2\over k_{\perp,min}^2 + K_z^2}\right]
\label{eq:eps2}
\end{equation}
where for the line lattice, $k_{\perp,max}\sim \pi/\xi_0$,
and for $\lambda\gg a_v$,
$k_{\perp,min}\sim \pi/a_v$.
For our data, we find a reasonable fit with
$k_{\perp,max}\approx\pi$ (in our model $\xi_0\equiv 1$),
but $k_{\perp,min}\approx 0$.

{}From the above analysis, we can extract several parameters of interest.
Considering the $(k_\perp$, $k_z)\to 0$ limits, we fit our data to the
form of Eq.(\ref{eq:tks}) and extract the values of the hydrodynamic
parameters $J_1$ and $\epsilon_1(0)$, which we plot versus $T$ in
Figs.15$a$ and $b$, respectively.
Defining the prefactor of the logarithm
in Eq.(\ref{eq:eps2}) as $(\pi/2)
J_2$ (as motivated by Eq.(\ref{eq:J})),
we can fit the large $K_z^2$
part of the data in Fig.14 to $\sim\ln(K_z^2)$ and plot the resulting
$J_2$ in Fig.15$a$.  We see that $J_1$ and $J_2$ are of
comparable magnitude,
except at the very lowest temperature
$T=0.6J_0$, and above $T_{cz}\simeq
2.5J_0$, where $J_2$ decays to zero, while $J_1$ remains approximately
constant.

Finally, we can extract a correlation length along the field direction,
by computing the Fourier transform of $S(k_\perp,k_z)$ with respect to
$k_z$, as in Eq.(\ref{eq:Skz}).  Fitting the resulting $S(k_\perp,z)$,
at large $z$, to the form
\begin{equation}
S(k_\perp,z)\simeq S_0(k_\perp)\left[ {\rm e}^{-z/\xi(k_\perp)}
  +{\rm e}^{-(L_z-z)/\xi(k_\perp)}\right]
\label{eq:S0}
\end{equation}
(this form being chosen so as to account for the periodic boundary
conditions along $\hat z$), we plot the correlation length
$\tilde\xi_z\equiv \xi(k_\perp=2\pi/a_v)$
versus $T$ in Fig.5.  For $T<2.25J_0$ we found no measurable
decay in $S(k_\perp,z)$.  For $T\ge 2.25J_0$ we find
$\tilde\xi_z\approx \xi_c$,
the cutting correlation length.  This reinforces our
earlier observation that
cutting and entangling appear to occur on the same length scales.
We can use the value of $\tilde\xi_z$ just obtained, to estimate
the hydrodynamic line tension.  For finite $\tilde\xi_z$, we
expect that the effective $k_z\to 0$ line tension can be
obtained from Eq.(\ref{eq:eps2}) (appropriate for the line lattice
where $\tilde\xi_z\to\infty$) by
imposing a small $k_z$ cutoff at $k_z\simeq\pi/\tilde\xi_z$.
Hence we estimate,
\begin{equation}
\epsilon_1(0)\simeq
\epsilon_1(k_z\simeq \pi/\tilde\xi_z)\simeq\pi J_2\ln \tilde\xi_z
\label{eq:eps3}
\end{equation}
We plot the result in Fig.15$b$, where we have
set $\tilde\xi_z\equiv L_z$ for all $T<T_{cz}$.  Although we observe
quite reasonable agreement with
the hydrodynamic result obtained from a direct fit to Eq.(\ref{eq:tks}),
we note
that the dominant temperature
dependence comes from the behavior of $J_2$.
We do not have large enough of a range of $\tilde\xi_z$ to provide a
sensitive test of the $\ln \tilde\xi_z$ dependence.

\noindent {\bf IV. Conclusions and Discussion}

To conclude, we have studied a
model of fluctuating vortex lines, and find
three distinct thermodynamic phases.  Upon heating, the system has sharp
transitions from a vortex line lattice, to an unentangled
vortex line liquid,
to an entangled vortex line liquid with many closed vortex
line loops and much
vortex line cutting.  The lower transition is associated
with the loss of phase
coherence in the superconducting order parameter, in
directions perpendicular
to the applied magnetic field.  The upper transition
is associated with the
loss of phase coherence in the parallel direction.
This behavior is consistent
with that argued by Fiegel'man and coworkers\cite{Fei}.
Similar behavior has been argued
by Glazman and Koshelev\cite{Glaz}, for low magnetic
fields $B<B_{cr}$\cite{note}.
Nelson\cite{Nel} originally suggested an unentangled line liquid
phase as a finite
size effect, when $\xi_z\approx L_z$.  We cannot rule out the
possibility that the
difference we observe between $T_{c\perp}$ and $T_{cz}$
is such a finite size
effect.  However, if this is so, our observation
that $T_{cz}\simeq 2T_{c\perp}$
would indicate a rather large critical region about
the single transition of
the infinite system.  Similarly, more detailed finite
size studies would be
needed for one to clarify the {\it nature} of the phase
transitions in our
model.  In our earlier work\cite{LiTei1,LiTei2}, at the
higher density $f=1/5$,
we found evidence for only a single transition. It may be
that there is indeed
only a single transition at higher densities, or it may be
that the width of
the intermediate region becomes too narrow to distinguish.
It is interesting
to note that in a related model, the $3d$XY model of stacked
triangular planes
with antiferromagnetic coupling within the triangular planes
and ferromagnetic
coupling between the planes, which has the same symmetry\cite{DHL}
as our model
with $f=1/2$, there is only a single transition, however it is a
tetracritical point\cite{f12}.

Our model has involved two key approximations in its mapping to a
realistic superconductor.
One is the discretization of space into a cubic
mesh, on the length scale of the bare coherence length $\xi_0$.
The discretization along the $\hat z$ axis may be viewed as representing
the layered structure of the
high $T_c$ materials, as in a Lawrence-Doniach
model\cite{LawDon}.  The discretization in the $xy$ plane
however, represents an artificial periodic pinning potential for
vortex line fluctuations.  Our hope was that this pinning potential
primarily acts to remove the mode associated with a uniform translation
of the entire vortex line lattice.
However the ``forces" associated with
this pinning potential are strong enough to tilt the balance in favor of
a square line lattice in the ground state, as opposed to the triangular
lattice expected in a continuum. One may worry that these forces
also significantly
suppress vortex line fluctuations, leading to a stabilization
of the line lattice, as well as phase coherence of the superconducting
order parameter, at low temperatures.
However, in Figs.15 we see a sharp
drop in $J_1$ and $\epsilon_1(0)$ at a $\tilde T< 1.0J_0$, noticeably
below the lower transition $T_{c\perp}\simeq 1.3J_0$.  If we interpret
$\tilde T$ as the energy scale
associated with the energy barriers of the periodic pinning potential,
then one
expects that the transition at $T_{c\perp}$ is a true melting transition
of the line lattice, rather than just a depinning of the lines.

The second approximation has been the assumption of a uniform magnetic
induction
$\vec B$ inside the superconductor, ie. the $\lambda\to\infty$ limit.
We first discuss the effects of this approximation on vortex line
fluctuations.
As mentioned in Section II, setting $\lambda\to\infty$ suppresses long
wavelength (small $k<\lambda^{-1}$) density fluctuations, while
being a good
approximation for high $k
\mathrel{\hbox{\hbox to 0pt
{\lower.5ex\hbox{$\sim$}\hss}\raise.4ex\hbox{$>$}}}
\lambda^{-1}$ fluctuations.
It is generally believed\cite{Hou,Glaz}
however, that it is high $k_\perp\simeq 2\pi/a_v$ shear fluctuations,
which do not alter
the long wavelength density, that are responsible for melting of the
vortex line lattice.  Ikeda {\it et al.}\cite{Ikeda} specifically
note that
the melting transition, as determined from a Lindemann criterion, is
largely unaffected when comparing the finite $\lambda$ to the
$\lambda\to\infty$
cases, providing $\lambda\gg a_v$.  Hence we expect behavior in the line
lattice phase, and the melting at $T_{c\perp}$,
to be largely independent
of this uniform $\vec B$ approximation.  Similarly, in the normal,
entangled
line liquid phase, $T>T_{cz}$, one expects this approximation to be good
for describing $local$ properties (such as entangling, cutting,
and $S(\vec k)$
for $k>\lambda^{-1}$) provided $\lambda\gg\tilde\xi_z$, the
correlation length along the $\hat z$ axis.  In this case,
fluctuations impose a shorter cutoff on effective interactions
than does magnetic screening.
It is less clear how valid the approximation is in the intermediate
phase, where $\tilde\xi_z\to\infty$,
and there is no other natural length
scale along the $\hat z$ direction with which to compare $\lambda$.
However we note that Feigel'man\cite{Fei} predicts a similar unentangled
phase, working
within the ``$2d$ boson" approximation, in which the interaction between
vortex lines along the $\hat z$ direction is short ranged.
Further work will be necessary to determine whether the unentangled line
liquid phase observed here, is stable against fluctuations of $\vec B$.

The effect of the $\lambda\to\infty$ approximation on fluctuations of
the phase of the superconducting order parameter,
and the existence of true long range order in the gauge
invariant order parameter correlation function, has been a subject
of much discussion\cite{Glaz,Ikeda,Moo2,Hou2}.
Moore\cite{Moo2} and
Ikeda {\it et al.}\cite{Ikeda} have shown that in three dimensions,
the gauge invariant order parameter
correlations decay algebraically, in the line lattice phase, for the
$\lambda\to\infty$ case.
However for the finite $\lambda$ case, they decay exponentially.
Such algebraic decay is consistent with a finite helicity
modulus\cite{note2}
$\Upsilon$, as we observe in our model.  Exponential decay,
for finite $\lambda$,
would imply $\Upsilon\to 0$.  However as shown in
Refs.\cite{Ikeda,Moo2},
the length scale for such exponential decay is given by
$(\pi J_0/T)\lambda\kappa$, and is hence comparable to the size
of macroscopic samples, for the large $\kappa$ high $T_c$ materials.
Thus the $\lambda\to\infty$ limit should be appropriate in such
situations\cite{note3}.

Although we have not attempted any direct mapping between our model and a
specific high $T_c$ material (we took isotropic couplings,
and ignored all $T$ dependence of our bare parameter $J_0$),
it is tempting to try to discuss resistivity measurements in very pure
YBCO single crystals\cite{Wor,Kwok2},
in terms of the behavior we observe in our model.
Above the upper transition $T_{cz}$, where there are many vortex line
loops, much wandering and entangling of the field induced vortex lines,
and easy cutting of vortex lines, we might expect substantial flux flow
resistance due to the flow of effectively $independent$ vortex
line segments.
This region possibly describes much of the broadening of the resistive
transition, in comparison to the $B=0$ case, and could be explained
with traditional flux flow models\cite{Tink2}.
Resistivity, $\rho_z$, for current $I\parallel B$,
should be comparable to $\rho_\perp$ for $I\perp B$ (apart from effects
due to explicit anisotropy of couplings),  as
there are almost as many vortex line segments oriented perpendicular
to $B$ as parallel to $B$ (from closed loops, and transverse segments
of wandering lines).  As $T$ is cooled towards $T_{cz}$ we would
expect $\rho_z$ to decrease substantially compared to $\rho_\perp$,
as the vortex lines disentangle,
and straighten parallel to $B$, and there
are fewer loop excitations.  Such behavior is
consistent with that seen in experiments\cite{Kwok2}
where $\vec B$ is applied in the
$ab$ (CuO) planes, and hence there is no
explicit anisotropy effects for $\rho_z$ versus $\rho_\perp$.
Below $T_{cz}$, the
linear resistivity $\rho_z$ should vanish, however
the linear part of $\rho_\perp$ would remain finite.
The transition at $T_{cz}$ might offer an explanation for the
``kink" observed in $\rho_\perp$
in the tail of the resistance curve, for
very pure samples with $\vec B$ along the $c$ axis\cite{Wor}.
Finally, $\rho_\perp$ would vanish below $T_{c\perp}$.

\noindent {\bf Acknowledgements}

This work was supported by
grant DE-FG02-89ER14017 from the U.S. Department of Energy.

\newpage
\noindent{\large \bf Figure Captions}

\begin{quote}
{\bf Figure 1.} Ground state vortex line lattice for a magnetic
induction of $f=B\xi_0^2/\Phi_0=1/25$ flux quantum per unit cell of
the numerical mesh.  The view is along the direction of $\vec B$,
and (+) locates the positions of the straight vortex lines.
\end{quote}

\begin{quote}
{\bf Figure 2.} Helicity modulus $\Upsilon_z$
along direction of $\vec B$,
and $\Upsilon_\perp$ perpendicular to $\vec B$,
for both heating and cooling.
The vanishing of $\Upsilon_{z,\perp}$ indicate two
separate phase transitions.
$R$ is the fraction of vortex lines which reconnect to
themselves, when the
system is periodically extended along the $\hat z$ direction
(see section IIIB($i$)).
Vanishing of $R$ indicates entanglement of lines.
\end{quote}

\begin{quote}
{\bf Figure 3.} Schematic, showing how vortex lines must
connect onto each other,
due to the periodic boundary conditions along the $\hat z$ axis.
Solid lines
represent vortex lines.  Dashed lines indicate how the ends of
the vortex lines must match up on the $z=0$ and $z=L_z$ planes.
\end{quote}

\begin{quote}
{\bf Figure 4.} Transverse fluctuation of vortex lines
$\langle u^2\rangle$ as
a function of length along the line $l_z$, for several
different temperatures.
Linear growth with $l_z$ indicates  wandering, entangled lines.
Saturation with increasing $l_z$ indicates unentangled lines.
\end{quote}

\begin{quote}
{\bf Figure 5.} Correlation lengths along the direction of $\vec B$.
$\xi_z$ is the
``entanglement"  length, determined by the transverse
fluctuations of Fig.4.
$\xi_c$ is the ``cutting" length,
giving the average distance between two
cuttings (intersections) along a
single vortex line.  $\tilde\xi_z$ is the
decay correlation length as extracted from the vortex line
structure function, $S(\vec k_\perp,z)$ (see section IIIC).
\end{quote}

\begin{quote}
{\bf Figure 6.} Average density of vortex lines $n_v$, piercing planes
of constant $z$.
At low $T$, $n_v$ is the magnetic field induced
density of $f=1/25$.  $n_v$
increases with $T$ due to the excitation of closed vortex line loops.
\end{quote}

\begin{quote}
{\bf Figure 7.} Specific heat $C$ versus $T$.
\end{quote}

\begin{quote}
{\bf Figure 8.} Schematic, showing the measurement of local bond
angles $\phi_{j,j+1}(i)$,
used for the calculation of local orientational order in
Eq.(\ref{eq:orient}).
\end{quote}

\begin{quote}
{\bf Figure 9.} 4-fold and 6-fold local orientational order parameters,
$\Psi_4=|\Psi_4|{\rm e}^{i\vartheta_4}$ and
$\Psi_6=|\Psi_4|{\rm e}^{i\vartheta_6}$.
The kink in $|\Psi_{4,6}|$ indicates $T_{c\perp}\simeq 1.3J_0$,
while the crossover
in $\vartheta_{4,6}$ indicates a transition from local
4-fold to 6-fold order.
\end{quote}

\begin{quote}
{\bf Figure 10.} Intensity plots of vortex structure function
$S(\vec k_\perp,z=0)$, for
various temperatures.
$k_x$ is the horizontal, and $k_y$ is the vertical
direction.  Brightness measures the magnitude of $S$, as in a
diffraction pattern.
The $\delta$-function peak at $\vec k_\perp=0$ is the temperature
independent constant $S(0,z=0)=f^2L_\perp^2$, which sets the relative
intensity scale.  ($a$)
shows a lattice of Bragg peaks, characteristic of a vortex line lattice,
at $T<T_{c\perp}$.  ($b$) and ($c$) are characteristic of a vortex line
liquid, for $T_{c\perp}<T<T_{cz}$.  ($d$) is for $T>T_{cz}$, and the
dominating continuous
background indicates the proliferation of closed vortex line loop
excitations.
\end{quote}

\begin{quote}
{\bf Figure 11.} Time decay of real space vortex correlation function
of Eq.(\ref{eq:gt})
for $T=1.9J_0$, in the intermediate phase.  The linear behavior on the
log-log
plot indicates algebraic decay with the power $0.23$.  The unit of
time is one Monte Carlo
sweep through the entire numerical mesh.  The solid line is a guide
to the eye.
\end{quote}

\begin{quote}
{\bf Figure 12.} Intensity plot of vortex structure function
$S(\vec k_\perp,z=0)$, for
$T=0.6J_0$, upon cooling the system through $T_{c\perp}$.
\end{quote}

\begin{quote}
{\bf Figure 13.} Structure function data, plotted as
$TK_\perp^2/S(k_\perp,k_z)$, versus
$K_\perp^2$ and $K_z^2$, for several different temperatures.
$K_\mu^2$ are
related to $k_\mu$ as in Eq.(\ref{eq:Green}).
The different curves versus $K_\perp^2$,
are for different values of $k_z=2\pi m_z/L_z$.
The different curves versus $K_z^2$, are
for different values of $k_\perp=2\pi m_\perp/L_\perp$
($m_\perp=10$ corresponds to $k_\perp=2\pi/a_v$).
($a$) and ($b$) are for
$T=1.0J_0$ in the line lattice phase; ($c$) and ($d$) are
for $T=2.25J_0$ in the
intermediate, unentangled line liquid phase; ($e$) and
($f$) are for $T=2.75J_0$ in
the normal, entangled line liquid phase.
\end{quote}

\begin{quote}
{\bf Figure 14.} Line tension $\epsilon_1(k_z)$, from
Eq.(\ref{eq:eps1}), using data
as in Fig.13.  The linear dependence on $\ln K_z^2$, at large $K_z^2$,
agrees qualitatively with the theoretical
prediction of Eq.(\ref{eq:eps2}).
$K_z^2$ is related to $k_z$ as in Eq.(\ref{eq:Green}).
\end{quote}

\begin{quote}
{\bf Figure 15.} ($a$) Coupling $J_1$, and ($b$) line tension
$\epsilon_1(k_z\to 0)$,
obtained from a fit of $TK_\perp^2/S$ to the hydrodynamic limit,
Eq.(\ref{eq:tks}).
($a$) Coupling $J_2$, obtained from a high
$K_z^2$ fit of the line tension
data of Fig.14, to the form Eq.(\ref{eq:eps2}).  ($b$)
Hydrodynamic line tension
$\pi J_2\ln \tilde\xi_z$ obtained from Eq.(\ref{eq:eps3}), using
structure function correlation length $\tilde\xi_z$ of Fig.5.
\end{quote}


\begin{thebibliography}{99}
\bibitem{note} However, according to their estimate of $B_{cr}$
  (see Eq.(5) of Ref.\cite{Glaz}), we believe
  our model lies in the region $B>B_{cr}$.
\bibitem{note3} In Ref.\cite{Ikeda} it is argued that for both the
 finite $\lambda$ and the $\lambda\to\infty$ cases, the order parameter
 correlations, both perpendicular and parallel to $\vec B$, become short
 ranged once the shear modulus $c_{66}\to 0$.
 This would appear to contradict the observation in our model,
 that $\Upsilon_z>0$ in the intermediate phase,
 where presumably $c_{66}=0$.
 Our results also contradict the prediction of
 Ref.\cite{Glaz}, that phase
 coherence parallel to $\vec B$ is lost at a temperature much below the
 melting transition of the vortex line lattice.
\end{thebibliography}
\end{document}